\documentstyle[12pt]{article}
 \makeatletter \oddsidemargin  -.1in \evensidemargin -.1in
\newcommand{\singlespacing}{\let\CS=\@currsize\renewcommand{\baselinestretch}{1}\tiny\CS}
\newcommand{\oneandahalfspacing}{\let\CS=\@currsize\renewcommand{\baselinestretch}{1.25}\tiny\CS}
\newcommand{\doublespacing}{\let\CS=\@currsize\renewcommand{\baselinestretch}{1.35}\tiny\CS}

\newtheorem{rule-def}[theorem]{Rule}

\RequirePackage[dvips]{graphicx} \textheight 21.5cm

\begin{document}

\title{\bf Solution of generalized magnetothermoelastic problem by using finite difference method}
\author{\small B. Das  \thanks{Corresponding Author:
bappa.das1@gmail.com (Bappa Das)} $^{}$ ,~\small G. C. Shit
\thanks{Email address: gopal\_iitkgp@yahoo.co.in (Gopal Chandra Shit) } ~ and A. Lahiri \thanks{Email address:
lahiriabhijit2000@yahoo.com (Abhijit Lahiri)
} \\
\it {$^*$}Department of Mathematics,\\
\it Ramakrishna Mission Vidyamandira, Belur Math,\\\it Howrah - 711202.\\
\it {{$^\dag$,}{$^\ddag$}}Department of Mathematics,\\
\it Jadavpur University,Kolkata - 700032. }
\date{}
\maketitle \noindent \doublespacing
\noindent {\bf Abstract -}
 A magnetothermoelastic problem is considered for a nonhomogeneous, isotropic rotating hollow cylinder in the context of three theories of generalized formulations, the classical dynamical coupled (C-D) theory, the Lord and Shulman's (L-S) theory with one relaxation time parameter as well as the Green and Lindsay's (G-L) theory with two relaxation time parameters. The inner surface of the cylinder is subjected to a time dependent exponential thermal shock at its inner boundary. The inner and outer surfaces of the hollow cylinder are assumed to be traction free and the temperature gradient vanishes at its outer surface. The problem is solved numerically using finite difference method by developing Crank-Nicolson implicit scheme. The numerical computations of the displacement component, temperature distribution, radial and hoop stresses have been estimated. A comparison has been made under the effect of different parameters by representing several figures.\\

\noindent {\bf Key Words -}
Finite Difference Method (FDM), Generalized Magnetothermoelasticity, Isotropic and Nonhomogeneous.\\

\section{Introduction}
A large number of research works carried out in the field of thermoelasticity (coupled or generalized). Biot \cite{R1} first introduced classical coupled thermoelasticity which predicts infinite speed of wave propagation in homogeneous and isotropic elastic medium. To make it more physical relevant, the theory of conventional coupled thermoelasticity was modified, Lord-Shulman \cite{R2} introduced 'generalized thermoelasticity' to ensure the propagation of thermoelastic wave in finite speed introducing one thermal relaxation time parameter to the energy equation without violating the conventional theory of Fourier's law of heat conduction. Not only the isotropic cases, Lord-Shulman (L-S) theory is equally applicable for the cases of anisotropic medium which was experimented theoretically by Dhaliwal and Sherief \cite{R3}. Also uniqueness of this theory was verified by many scholars as Ignaczak \cite{R4}, Sherief \cite{R5}. Later, Green-Lindsay (G-L model) \cite{R6} introduced another theory named 'second sound' which is generalized one just introducing two ralaxation time parameters, one in equation of motion and another in also heat conduction equation. Ignaczak and Starzewski \cite{R7} studied the thermoelasticity with finite wave speeds. Lahiri and Das \cite{R8} also worked on the problem of generalized thermomechanical interactions for an unbounded body with a circular cylindrical hole without energy dissipation.\\
Now-a-days, increasing attention is being devoted to the field of magnetothermoelasticity which is the interactions in between magnetic field and thermomechanical coupling of the isotropic and anisotropic medium. In nuclear reactors, high energy and high energy gradient fields also in geophysics, plasma physics are the fields where this theory has great applicability. Many researchers enriched this field of magnetothermoelasticity. Abd-Alla and Abo-Dahab \cite{R9} studied the generalized magnetothermoelastic effect in the influence of normal point loading and thermal point loading in the viscoelastic medium and Allam et.al. \cite{R10} also experimented theoretically a one dimensional magnetithermoelastic problem in the context of Green-Naghdi \cite{R11} model.\\
Most of thermoelastic problems (classical coupled or generalized) have been solved by (i) potential function approach (ii) state-space approach, and (iii) eigenvalue approach. We have solved the magnetothermoelastic problem using finite difference method (FDM) by developing Crank-Nicolson scheme.\\
The FDM is a numerical technique that finds an approximate solution of a given problem. The concept of this method is replacing the derivatives that appear in the differential equation by an algebraic approximation. The unknowns of the approximated algebraic equations are the dependent variables at the grid points.\\
Analytical solutions of PDEs provide us with closed -form expressions which depict the variations of the dependent variables in the domain. The numerical solutions, based on finite differences, provide us with the values at discrete points in the domain which are known as grid points.\\
In some class of problems, the numerical calculations are performed on a transformed computational plane which has uniform spacing in the transformed-independent-variables but non-uniform spacing in the physical plane.\\
In this article, we consider a thermoelastic problem for a non-homogeneous isotropic hollow cylinder subjected to a time dependent exponential thermal shock.
Finite difference method using Crank-Nicolson scheme is used to the basic equations of this problem for solving numerically. Finally, numerical computations of the displacement component, temperature distribution, and stress components have been done and represented graphically.\\

{\small{\bf {Nomenclature}}}\\

$\lambda,\mu $= Lam$\grave{e}$  constants.\\
$u_i$ = Displacement component.\\
$T $= Absolute temperature.\\
$T_0$ = Reference temperature chosen such that
$|\frac{T-T_0}{T_0}|<<1$.\\
$K_{ij}$ = Coefficient of thermal conductivity. $(i,j=1,2,3)$\\
$\mu_{0}$ = Magnetic permeability.\\
$h_{i}$ = Magnetic perturbation.\\
$\rho$ = Density of the medium.\\
$\gamma$ = Material Constant = $({3\lambda+2\mu}){\alpha_T}$\\
$\alpha_T$ = Coefficient of linear thermal expansion.\\
$t$ = Time variable.\\
$C_v$ = Specific heat of the material at constant strain.\\
$\tau _0$, $\tau _1$ = Thermal relaxation time parameters.\\
$\delta _{ij}$ = Kronecker delta.\\

\section{Basic Equations}
 We consider a non-homogeneous isotropic rotating hollow cylinder with imposing a time dependent exponential thermal shock. For thermoelastic problem,
 the equation of motion is given by
\begin{eqnarray}
\sigma_{ij,j} +\tau_{ij,j} = \rho[{\bf{\ddot{u}}+\{{\bf{\Omega}}\times({{\bf{\Omega}}\times{{\bf u}}})\}+
({2{{\bf {\Omega}}\times\dot{{{\bf {u}}}}}})}]_i,
\end{eqnarray}
The elastic medium is rotating uniformly with an angular velocity $\bf \Omega = \Omega\hat{n}$, where $\hat{\bf{n}}$ is a unit vector representing the direction of the axis of rotation. The displacement equation of motion in the rotating frame has two additional terms, viz., $\{\bf{\Omega} \times({{\bf{\Omega}}\times{{\bf u}}})\}$ is the centripetal acceleration due to time varying motion only and $({2{{\bf {\Omega}}\times\dot{{{\bf u}}}}})$ is the Coriolis acceleration with ${\bf{\Omega}}=(0,\Omega,0)$.\\
The constitutive stress tensor $\sigma_{ij}$ of Duhamel-Neumann form is given by
\begin{eqnarray}
\sigma_{ij} = \lambda u_{i,i} \delta_{ij} +\mu(u_{i,j}+ u_{j,i})-\gamma(T + \tau_1\dot{T})\delta _{ij}
\end{eqnarray}
and the Maxwell stress tensor $\tau_{ij}$ is given by
\begin{eqnarray}
\tau_{ij} = \mu_{0}(h_iH_j+h_jH_i-h_kH_k \delta_{ij})
\end{eqnarray}
The constitutive strain tensor $e_{ij}$ in terms of displacement components $u_i$ has the form
\begin{eqnarray}
e_{ij} = \frac{1}{2}(u_{i,j}+u_{j,i})
\end{eqnarray}
The heat conduction equation for this problem can be written as
\begin{eqnarray}
(K_{ij}T_{,j})_{,i}= \rho C_v(\dot{T}+\tau_0\ddot{T})+(1+n\tau_0\frac{\partial}{\partial t})\gamma T_0\dot{u}_{i,i}
\end{eqnarray}
Subscript indices following a comma denote the partial derivatives with reference to  the cylindrical coordinate system while the superposed dot represents the time derivative.\\
The three equations of classical dynamics(C-D) coupled theory, Lord and Shulman (L-S) theory and Green and Lindsay
(G-L) theory as follows \\
(i) Classical dynamical coupled theory (C-D, 1956)\\
  ~~~~~~~~~~~~~~~~~~~~   $\tau_0$=0,~~~ $\tau_1$=0, ~~~$n=0$\\
  (ii) Lord and Shulman's theory (L-S, 1967)\\
  ~~~~~~~~~~~~~~~~~~~~   $\tau_0$$>$0,~~~ $\tau_1$=0, ~~~$n=1$\\
  (ii) Green and Lindsay's theory  (G-L, 1972)\\
  ~~~~~~~~~~~~~~~~~~~~   $\tau_0$$>$0,~~~ $\tau_1$$>$0, ~~~$n=0$ .\\
For generalized thermo-elastic body, the field equations and electromagnetic quantities satisfy Maxwell's equations may be written as
\begin{eqnarray}
\textrm{curl} ~{\bf{h}} = {\bf{J}} + \varepsilon_0\frac{\partial {\bf{E}}}{\partial t}~,~
\textrm{curl} ~{\bf{E}} = - \mu_0\frac{\partial {\bf{h}}}{\partial t}\nonumber\\
\textrm{div} ~ {\bf{h}} = 0,~~~~~~~~{\bf{E}} = -\mu_0(\dot{\bf{u}}\times \bf{H})\nonumber\\
{\bf{B}} = \mu_0({\bf{H}}+ {\bf{h}}),~~~~~~~~{\bf{D}} = \varepsilon_0{\bf{E}}
\end{eqnarray}

For nonhomogeneous characteristics of the material, we have
\begin{eqnarray}
(\lambda,\mu, \mu_0, \rho, K, ~\gamma)= f(\overrightarrow{r})(\lambda^*,\mu^*, \mu_0^*,\rho^*,K^*,\gamma^*)\nonumber\\
=(\lambda^1,\mu^1, \mu_0^1,\rho^1,K^1,\gamma^1)
\end{eqnarray}
where $\lambda^*$, $\mu^*$, $\rho^*$, $K^*$ and $\gamma^*$  are constants values of $\lambda$, $\mu$, $\rho$, $K$ and $\gamma$ respectively in the homogeneous case and $f(\overrightarrow{r})$ denotes the  non-dimensional continuous functions.\\

%
%
%

\section{Problem Formulation}
 Let us now consider an axially symmetric nonhomogeneous, isotropic and thermo-elastic hollow cylinder of inner and outer radii '$a_1$' and '$a_2$' respectively with cylindrical polar coordinates $ (r,\theta, z)$ under the action of magnetic field of strength $H_0$. Then the total magnetic field becomes ${\bf{H}}={\bf{H_0}}+{\bf{h}}$, where ${\bf{H_0}}=(0,0,H_0)$ is the initial magnetic field acting in the direction of the z-axis and also under the induced electric field ${\bf{E}}$. We also assume that both ${\bf{h}}$ and ${\bf{E}}$ are small in magnitude in accordance with the assumption of the linear theory of thermoelasticity.\\
 It is also considered that there is no body forces acting inside the medium and all thermal and mechanical interactions depend only on the radial distance $r$.\\
 Thus only the radial displacement exists and is given by
\begin{eqnarray}
u_r = u_r(r,z,t)= u
\end{eqnarray}
while the other form of displacements are all zero.
Similarly, the strain components $e_{rr}$ and $e_{\theta\theta}$ are exist and given by
\begin{eqnarray}
 e_{rr}=\frac{\partial u}{\partial r}~,~
 e_{\theta \theta} = \frac{ u}{ r}. 
\end{eqnarray}

From the equations (2) and (3), the stress components become
\begin{eqnarray}
\sigma_{rr} = 2\mu \frac{\partial u}{\partial r}+ \lambda (\frac{\partial u}{\partial r}+\frac{u}{r}) - \gamma(T + \tau_1\frac{\partial T}{\partial t}), \nonumber\\
\sigma_{\theta \theta}= 2\mu \frac{u}{r}+ \lambda (\frac{\partial u}{\partial r}+\frac{ u}{r}) - \gamma(T + \tau_1\frac{\partial T}{\partial t}), \nonumber\\
\sigma_{zz}= \lambda (\frac{\partial u}{\partial r}+\frac{u}{r}) - \gamma(T + \tau_1\frac{\partial T}{\partial t}), \nonumber\\
\tau_{rr}= \mu_0H^2_0 (\frac{\partial u}{\partial r}+\frac{u}{r}).
\end{eqnarray}
It is assumed that the material properties depend only on the radial coordinate $r$ and hence we write in equation (7) $f(\overrightarrow{r})=f({r})$. In the absence of body force and inner heat source, use of equations (1) and (5), we get the equation of motion as
\begin{eqnarray}
\frac{\partial \sigma_{rr}}{\partial r}+\frac{\sigma_{rr}-\sigma_{\theta \theta}}{r}+\frac{\partial \tau_{rr}}{\partial r}=\rho [\frac{\partial^2 u }{\partial t^2}-\Omega^2 u-2\Omega \frac{\partial u}{\partial t}]
\end{eqnarray}
and the heat conduction equation becomes
\begin{eqnarray}
\frac{1}{r}\frac{\partial}{\partial r}(r K \frac{\partial T}{\partial r})=\rho c_v(\frac{\partial}{\partial t}+\tau_0\frac{\partial^2}{\partial t^2})T+\gamma T_0(\frac{\partial}{\partial t}+n \tau_0\frac{\partial^2}{\partial t^2})(\frac{\partial u}{\partial r}+\frac{ u}{r}).
\end{eqnarray}
Using equation (7) for nonhomogeneous case, the equations (10)-(12) become (by dropping the symbol as one in superscript position, for convenience) the same form as of equation (10).

The equation of motion (11) becomes
\begin{eqnarray}
f(r)[H_a(\frac{\partial^2u}{\partial r^2}+\frac{1}{r}\frac{\partial u}{\partial r}-\frac{u}{r^2})-\gamma \frac{\partial}{\partial r}(T+\tau_1\frac{\partial T}{\partial t})-\rho (\frac{\partial^2 u }{\partial t^2}-\Omega^2 u-2\Omega \frac{\partial u}{\partial t})]\nonumber\\=-\frac{\partial f(r)}{\partial r}[H_a \frac{\partial u}{\partial r}+(H_a-2\mu)\frac{u}{r}-\gamma(T+\tau_1\frac{\partial T}{\partial t})]
\end{eqnarray}
and the heat conduction equation (12) gives rise to
\begin{eqnarray}
f(r)[\frac{\partial^2 T}{\partial r^2}+\frac{1}{r}\frac{\partial T}{\partial r}]=- \frac{\partial f(r)}{\partial r} \frac{\partial T}{\partial r}+ f(r)\frac{\rho c_v}{K}(\frac{\partial}{\partial t}+\tau_0\frac{\partial^2}{\partial t^2})T           \nonumber\\+f(r)\frac{\gamma T_0}{K}(\frac{\partial}{\partial t}+n \tau_0\frac{\partial^2}{\partial t^2})(\frac{\partial u}{\partial r}+\frac{ u}{r})
\end{eqnarray}

Let us introduce the following non-dimensional quantities
\begin{eqnarray}
T'=\frac{T}{T_0}~,~(r' ,u')=\frac{c}{\chi} (r,u)~,~(t' ,\tau_0',\tau_1')=\frac{c^2}{\chi} (t,\tau_0,\tau_1)~,\nonumber\\
~(\sigma'_{rr},\sigma'_{\theta\theta},\sigma'_{zz},\tau'_{rr})=\frac{1}{\mu}(\sigma_{rr},\sigma_{\theta\theta},
\sigma_{zz},\tau_{rr}),\Omega'=\frac{\Omega \chi}{c^2}
\end{eqnarray}
With the help of equation (15), the stress components have the form

\begin{eqnarray}
\sigma_{rr} =r^{2m}[\xi^2 \frac{\partial u}{\partial r}+ (\xi^2-2)\frac{ u}{r}-\eta(T + \tau_1\frac{\partial T}{\partial t})], \nonumber\\
\sigma_{\theta \theta}= r^{2m}[(\xi^2-2) \frac{\partial u}{\partial r}+ \xi^2\frac{ u}{r}-\eta(T + \tau_1\frac{\partial T}{\partial t})], \nonumber\\
\sigma_{zz}= r^{2m}[(\xi^2-2) (\frac{\partial u}{\partial r}+ \frac{ u}{r})-\eta(T + \tau_1\frac{\partial T}{\partial t})], \nonumber\\
\tau_{rr}= r^{2m}\omega^2(\frac{\partial u}{\partial r}+\frac{u}{r}).
\end{eqnarray}
The equation of motion (13) becomes
\begin{eqnarray}
\frac{\partial^2 u }{\partial t^2}-\Omega^2 u-2\Omega \frac{\partial u}{\partial t}=
\frac{\partial^2u}{\partial r^2}\nonumber\\+\frac{2m+1}{r}\frac{\partial u}{\partial r}+g\frac{u}{r^2}-\beta(1+\tau_1\frac{\partial }{\partial t})(\frac{2m}{r}T+\frac{\partial T}{\partial r})
\end{eqnarray}
and heat conduction equation (14) transforms to
\begin{eqnarray}
\frac{\partial^2 T}{\partial r^2}+\frac{2m+1}{r}\frac{\partial T}{\partial r}=(\frac{\partial}{\partial t}+\tau_0\frac{\partial^2}{\partial t^2})T+\varepsilon(\frac{\partial}{\partial t}+n \tau_0\frac{\partial^2}{\partial t^2})(\frac{\partial u}{\partial r}+\frac{ u}{r}),
\end{eqnarray}

where, $\xi^2=\frac{\lambda+2\mu}{\mu}$~;~$\omega^2=\frac{\mu_0H_0^2}{\mu}$~;~$\eta=\frac{T_0\gamma}{\mu}$~;~
$g=2m(1-\frac{2\mu}{H_a})-1$~;~$\beta=\frac{T_0\gamma}{H_a}$~;~$H_a=\lambda+2\mu+\mu_0H_0^2$~;~$\varepsilon=\frac{\gamma}{\rho c_v}$~;~
$c^2=\frac{H_a}{\rho}$~;~$\chi=\frac{K}{\rho c_v}$~;~$f(r)=(\frac{r}{\chi/c})^{2m}$~and $m$ taken as a non-dimensional rational number.\\

For the present problem, the appropriate initial and boundary conditions are taken as
\begin{eqnarray}
u(r,0)= \frac{\partial u(r,0)}{\partial t}=0~~;~~T(r,0)= \frac{\partial T(r,0)}{\partial t}=0
\end{eqnarray}
The inner and outer surface of the cylinder is assumed to be traction free. The temperature gradient at the external surface is also zero but a time dependent exponential thermal shock is applied to the internal surface of the cylinder, i.e.,
\begin{eqnarray}
\sigma_{rr}(a_1,t)=\sigma_{rr}(a_2,t)=0~~;~~\frac{\partial T(a_2,t)}{\partial r}=0~~;~~T(a_1,0)=e^{-kt}, k \mbox{ is constant}.
\end{eqnarray}

\section{Computational Scheme}
Using Crank-Nicolson scheme, finite difference discretization of
the governing equations (16)-(18) with appropriate initial and
boundary conditions have been carried out. The derivatives are
replaced by the central difference and time derivatives as forward
difference. We take the sub-divided rectangular domain into a
network by drawing straight lines parallel to the co-ordinate
axes. The solutions of $u(r,t)$ and $T(r,t)$ at any mesh point
$(i,j)$ in a computational domain are symbolized as $u_{i,j}^n$
and $T_{i,j}^n$ respectively in which $r_i=i\Delta r$,
$i=0,1,.....,m$ and $t_k=k\Delta t$, $k=0,1,.....$ where $m$ is
the maximum number of mesh points in the $r$-direction.\\
The discretized form of equations (17) and (18) are given by the following tri-diagonal system,
\begin{eqnarray}
[-\frac{1}{(\Delta r)^2}+\frac{2m+1}{2\Delta r_ir}]u_{i-1}^{n+1}+
[\frac{1}{(\Delta t)^2}-\Omega^2-\frac{2\Omega}{\Delta
t}+\frac{2}{(\Delta r)^2}-\frac{g}{r_i^2}]u_{i}^{n+1}\nonumber\\
+[-\frac{1}{(\Delta r)^2}-\frac{2m+1}{2\Delta r_ir}]u_{i+1}^{n+1}
=\frac{2u_i^n}{(\Delta t)^2}-\frac{u_i^{n-1}}{(\Delta
t)^2}-\frac{2\Omega}{\Delta
t}u_i^n-\frac{2m\beta}{r_i}[T_i^{n+1}+\tau_1\frac{T_i^{n+1}-T_i^n}{\Delta
t}]\nonumber\\
-\beta[\frac{T_{i+1}^{n}-T_{i-1}^n}{2\Delta
r}+\tau_1\frac{T_{i+1}^{n+1}-T_{i-1}^{n+1}-T_{i+1}^{n-1}+T_{i-1}^{n-1}}{4\Delta
r\Delta t}]
\end{eqnarray}
and
\begin{eqnarray}
[\frac{1}{(\Delta r)^2}-\frac{2m+1}{2\Delta r_ir}]T_{i-1}^{n+1}+
[-\frac{2}{(\Delta r)^2}-\frac{1}{\Delta t}-\frac{\tau_0}{(\Delta
t)^2}]T_{i}^{n+1}\nonumber\\
+[\frac{1}{(\Delta r)^2}+\frac{2m+1}{2 r_i\Delta r}]T_{i+1}^{n+1}
=-\frac{T_i^n}{\Delta t}+\tau_0\frac{-2T_i^{n}+T_i^{n-1}}{(\Delta
t)^2}+\varepsilon[\frac{u_{i+1}^{n}-u_{i-1}^{n}-u_{i+1}^{n-1}+u_{i-1}^{n-1}}{4\Delta
r\Delta
t}]+\nonumber\\
\frac{\varepsilon}{r}[\frac{u_{i}^{n+1}-u_{i}^{n-1}}{2\Delta
t}]+\varepsilon n
\tau_0[\frac{u_{i+1}^{n+1}-2u_{i+1}^n+u_{i+1}^{n-1}-u_{i-1}^{n+1}+2u_{i-1}^{n}-u_{i-1}^{n-1}}{2\Delta
r(\Delta t)^2}]+\nonumber\\
\frac{\varepsilon n
\tau_0}{r}[\frac{u_{i}^{n+1}-2u_{i}^{n}+u_{i}^{n-1}}{(\Delta
t)^2}]
\end{eqnarray}
After having determined $u$ and $T$ numerically, we can compute the stress
components which have the discretized form of equation (16) as follows

\begin{eqnarray}
\sigma_{rr} =r_i^{2m}[\xi^2 \frac{u_{i+1}^{n+1}-u_i^{n+1}}{\Delta r}+ (\xi^2-2)\frac{ u_i^{n+1}}{r_i}-\eta(T_i^{n+1} + \tau_1\frac{T_i^{n+1}-T_i^{n-1}}{2\Delta t})], \nonumber\\
\sigma_{\theta \theta}= r_i^{2m}[(\xi^2-2)\frac{u_{i+1}^{n+1}-u_i^{n+1}}{\Delta r}+\xi^2\frac{ u_i^{n+1}}{r_i}-\eta(T_i^{n+1} + \tau_1\frac{T_i^{n+1}-T_i^{n-1}}{2\Delta t})], \nonumber\\
\sigma_{zz}= r_i^{2m}[(\xi^2-2) (\frac{u_{i+1}^{n+1}-u_i^{n+1}}{\Delta r}+ \frac{ u_i^{n+1}}{r_i})-\eta(T_i^{n+1} + \tau_1\frac{T_i^{n+1}-T_i^{n-1}}{2\Delta t})], \nonumber\\
\tau_{rr}= r_i^{2m}\omega^2[(\frac{u_{i+1}^{n+1}-u_i^{n+1}}{\Delta
r}+ \frac{ u_i^{n+1}}{r_i})].
\end{eqnarray}

\section{Results and Discussion}
To illustrate the problem and comparing the theoretical results in the context of C-D, L-S and G-L theories, we have estimated some numerical results. For the purpose of numerical computation, we have consider the material to be copper (treated as isotropic), gives rise to the following data (cf. Dhaliwal and Singh \cite{R12}).\\
$\lambda  = 7.76\times10^{10}(Kg)(M)^{-1}(S)^{-2}$;~~$\mu  = 3.86\times10^{10}(Kg)(M)^{-1}(S)^{-2}$;~~
$\alpha_t  = 1.78\times10^{-5}(K)^{-1}$;~~
$c_v=383.1(M)^2(K)^{-1}(S)^{-2}$;~~$\mu_0  = 1(Gauss)(Oersted)^{-1}$;~~$K= 3.86\times10^{2}(Kg)(M)(K)^{-1}(S)^{-3}$;~~
$\rho = 8954(Kg)(M)^{-3}$.\\
The values of the associated constants used in this problem are taken as \\
$T_{0}  = 293^{0}(K);$~~$t=51(S)$;~~$H_0 = 10^3$;~~$m = -0.5,0.0,0.5$;~~$\tau_0, \tau_1=0.05(S)$;~~$n=0$;~~$\Omega  = 0,2.0,4.0~(Rad)(S)^{-1}$;~~${\it{a}_1}=1(M)$;~~${\it{a}_2}=2(M)$.\\

Figs. 1-14 exhibit the distribution of displacement ($u$), temperature ($T$) and stress component ($\sigma_{rr}$) verses $r$ for three fixed values of $m=-0.5,0,0.5$, $\Omega=0,2,4$ and $t=51$ for three theories C-D, L-S and G-L.

\textheight 22.0cm 
\begin{minipage}{1.0\textwidth}
  \begin{center}
\includegraphics[width=4.8in,height=3.5in ]{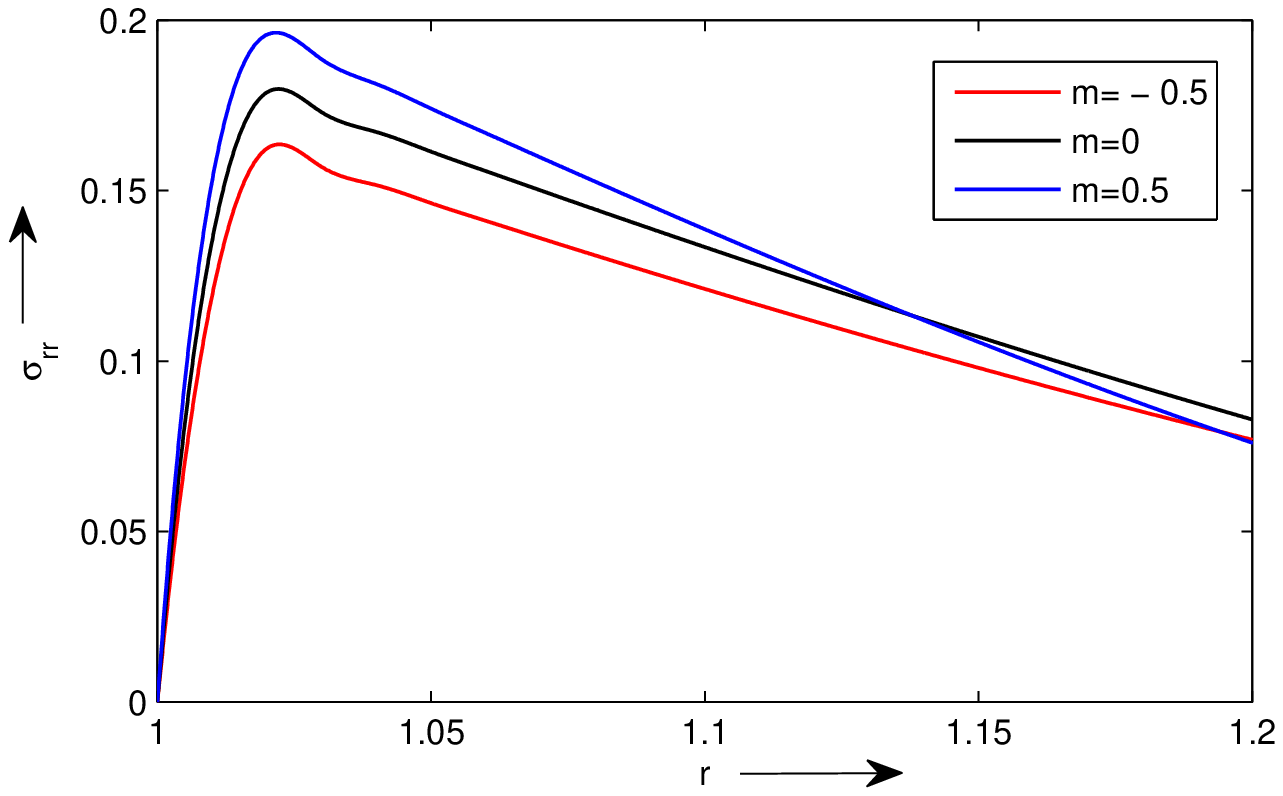}\\
Fig.1~~~ Distribution of stress ($\sigma_{rr}$) along with $r$ for $\Omega=0$ and fixed values of $m$ in G-L theory\\
\vspace*{1.5cm}
\includegraphics[width=4.8in,height=3.5in ]{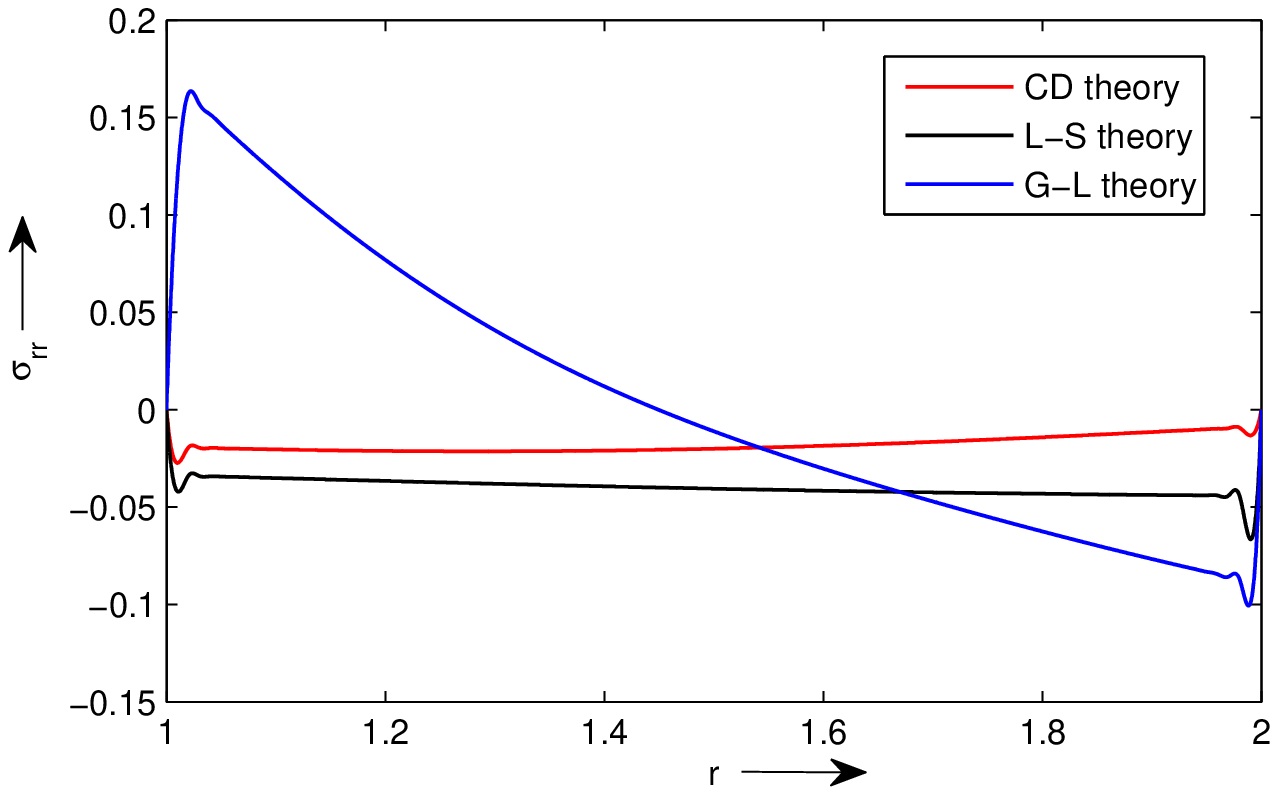}\\
Fig.2~~~ Distribution of stress ($\sigma_{rr}$) with $r$ for $\Omega=0$ and $m = - 0.5$ in three theories\\
\end{center}
\end{minipage}
\vspace*{.5cm}\\

\textheight 22.0cm 
\begin{minipage}{1.0\textwidth}
  \begin{center}
\includegraphics[width=4.8in,height=3.5in ]{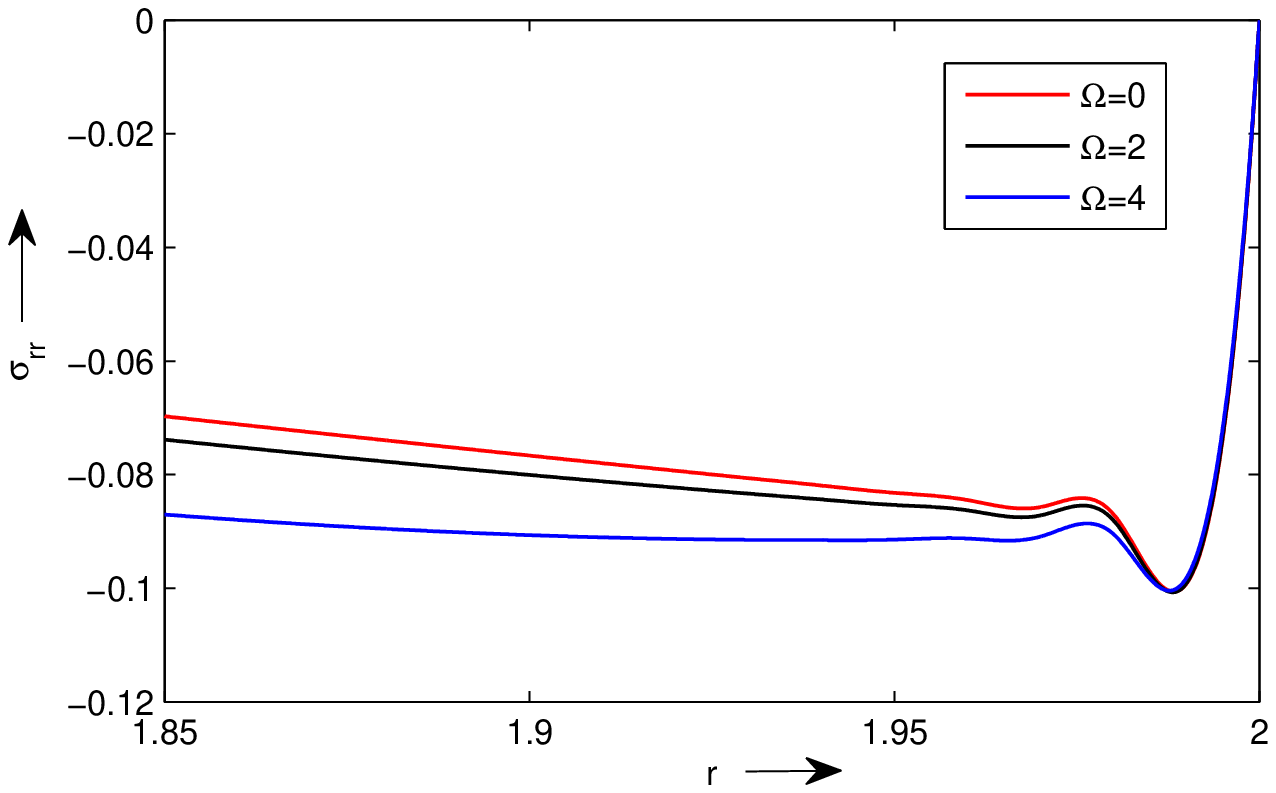}\\
Fig.3~~~ Distribution of stress ($\sigma_{rr}$) with $r$ for $m = - 0.5$ and fixed values of $\Omega$ in G-L theory\\
\end{center}
\end{minipage}
\vspace*{.5cm}\\

Figs. 1-3 give the distribution of radial stress component($\sigma_{rr}$)
 verses $r$ for different values of $m$ and $\Omega$. We observe that Fig.1 depicts the variation of $\sigma_{rr}$ in function of $r$ for three different values of $m=-0.5,0,0.5$ and $\Omega=0$ in G-L theory also in the region of
 $1\leq r\leq 1.2$. The absolute value of $\sigma_{rr}$ sharply attains the maximum at $r=1.02$ and it increases with the increase of $m$. In the region $1.02\leq r\leq 1.2$, the absolute value of $\sigma_{rr}$ attains the minimum value at $r=1.2$. Fig. 2 shows the variation of $\sigma_{rr}$ in three theories viz., C-D, L-S and G-L. The distributions of $\sigma_{rr}$ in C-D and L-S theory are almost parallel and in G-L theory, the variation of $\sigma_{rr}$ plays a significant characteristics. The absolute value attains the maximum in the neighboring points of $r=1$ and minimum in the neighboring point of $r=2$. Fig. 3 also gives the distribution of $\sigma_{rr}$ for $m=-0.5$ and in three fixed values of $\Omega=0,2,4$ in G-L theory. In the region $1.85\leq r\leq 1.96$, the absolute value of $\sigma_{rr}$ decreases very slowly as $r$ increases. In the region $1.97\leq r\leq 2.0$, the graphs of $\sigma_{rr}$ for $\Omega=0$, $\Omega=2$ and $\Omega=4$ coincide to each other. The absolute value of $\sigma_{rr}$ attains the minimum at $r=1.98$ and the maximum at $r=2.0$.\\

\textheight 22.0cm 
\begin{minipage}{1.0\textwidth}
  \begin{center}
\includegraphics[width=4.8in,height=3.5in ]{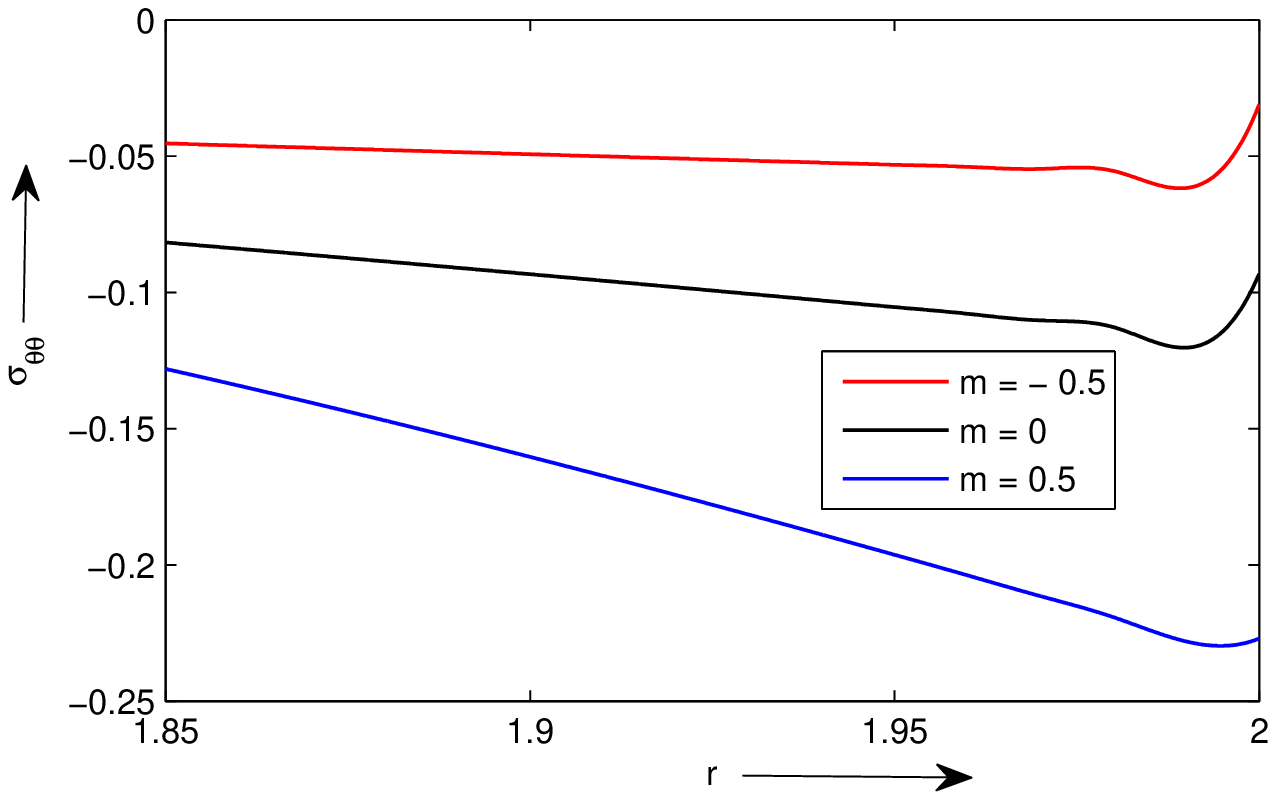}\\
Fig.4~~~ Distribution of stress ($\sigma_{\theta \theta}$) vs. $r$ for $\Omega=0$ and fixed values of $m$ in G-L theory\\
\vspace*{1.5cm}
\includegraphics[width=4.8in,height=3.5in ]{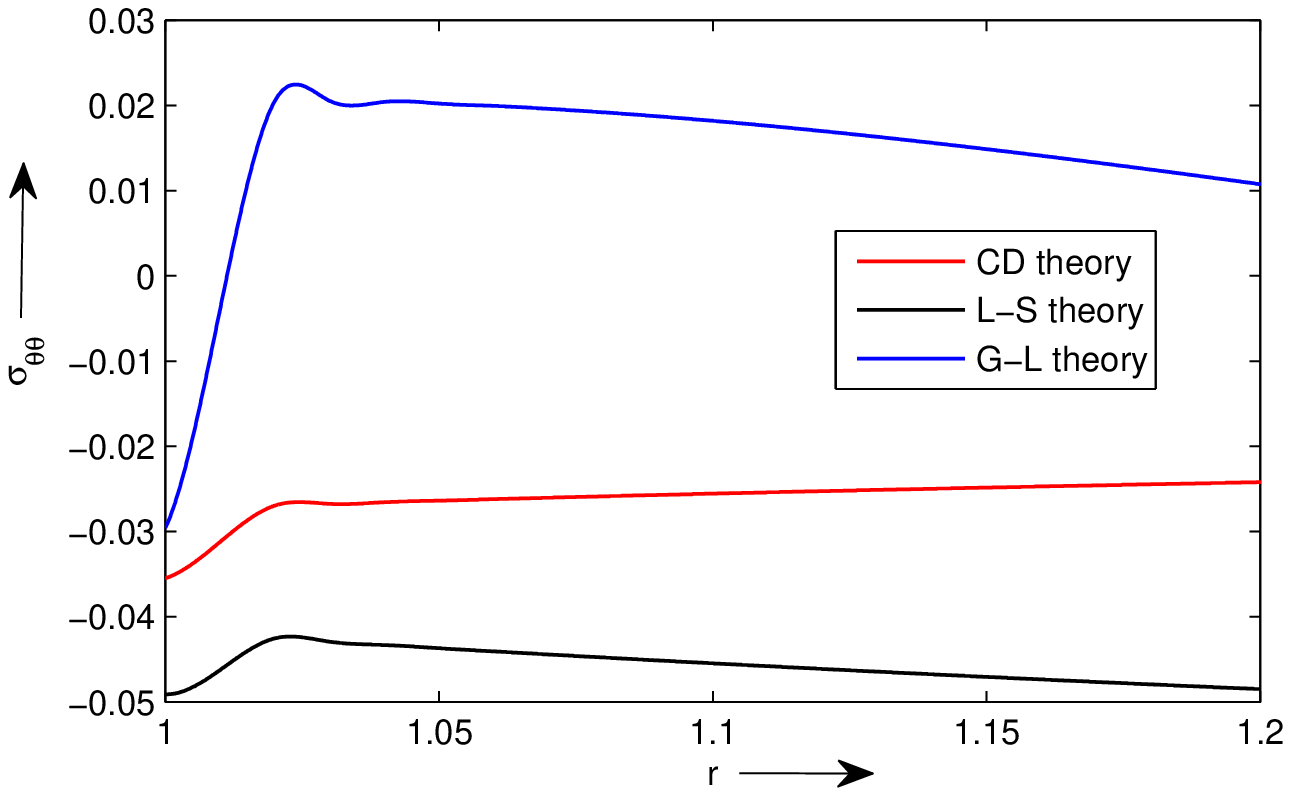}\\
Fig.5~~~ Distribution of stress ($\sigma_{\theta \theta}$) vs. $r$ for $m = - 0.5$ and $\Omega=0$ in three theories\\
\end{center}
\end{minipage}
\vspace*{.5cm}\\

\textheight 22.0cm 
\begin{minipage}{1.0\textwidth}
  \begin{center}
\includegraphics[width=4.8in,height=3.5in ]{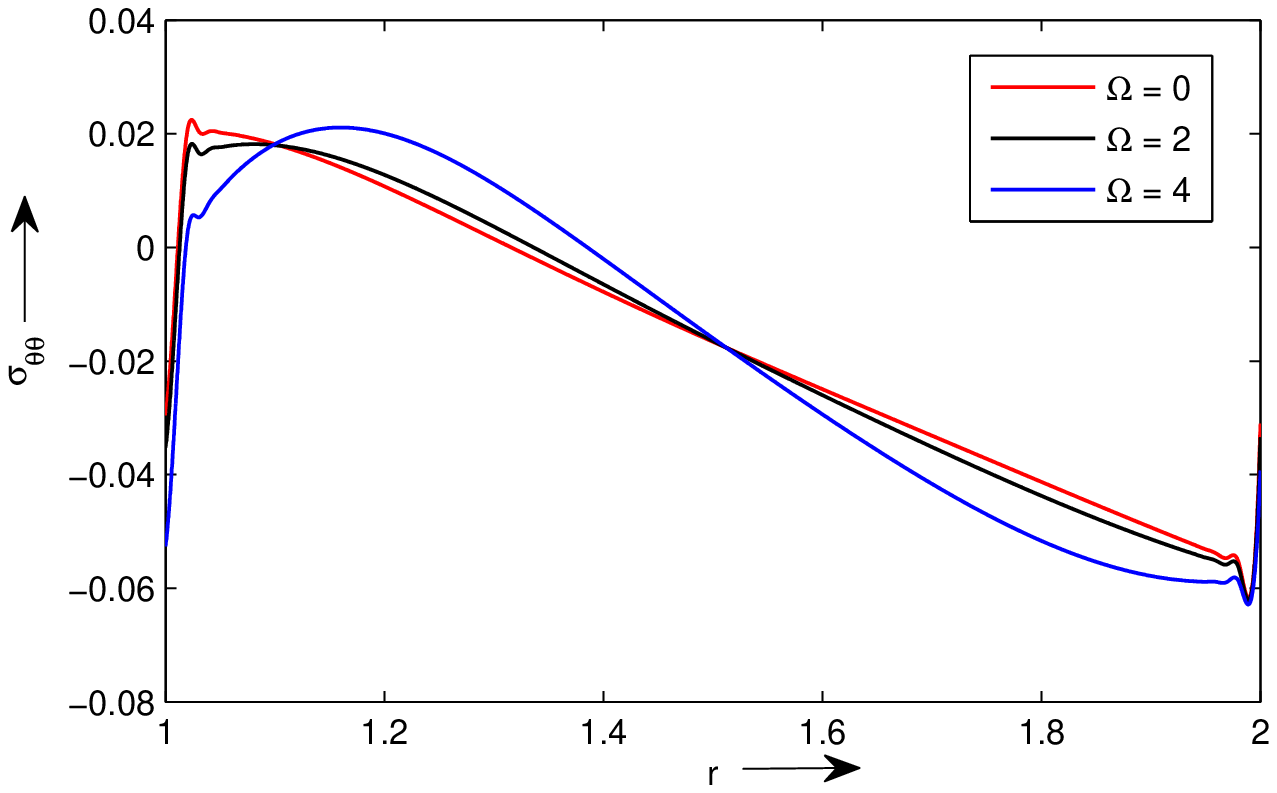}\\
Fig.6~~~ Distribution of stress ($\sigma_{\theta \theta}$) in function of $r$ for $m = - 0.5$ and fixed values of $\Omega$ in G-L theory\\
\end{center}
\end{minipage}
\vspace*{.5cm}\\

 Figs. 4-6 illustrate the distribution of $\sigma_{\theta\theta}$ along with $r$ for different values of parameters used in the problem. It is noticed from Fig.4 that the variation of $\sigma_{\theta\theta}$ for $\Omega=0$ and three fixed values of $m=-0.5,0,0.5$ in G-L theory. The absolute value of $\sigma_{\theta\theta}$ increases as $m$ increases. The absolute value of $\sigma_{\theta\theta}$ attains the maximum for $m=0.5$ in the neighboring point of $r=2$.
It is clearly observed from Fig. 5 that the graph of $\sigma_{\theta\theta}$ increases rapidly for three theories viz., C-D, L-S and G-L in the region $1\leq r\leq 1.02$. For G-L theory, the absolute value of $\sigma_{\theta\theta}$ decreases slowly as increment of $r$.
 Fig.6 shows the variation of $\sigma_{\theta\theta}$ for $m=-0.5$ and three fixed values of $\Omega=0,2,4$ in G-L theory. The absolute value of $\sigma_{\theta\theta}$ for three values of $\Omega=0,2,4$ sharply attains the maximum at $0.02$ and also rapidly decreases to the minimum value at $-0.06$.

\textheight 22.0cm 
\begin{minipage}{1.0\textwidth}
  \begin{center}
\includegraphics[width=4.8in,height=3.5in ]{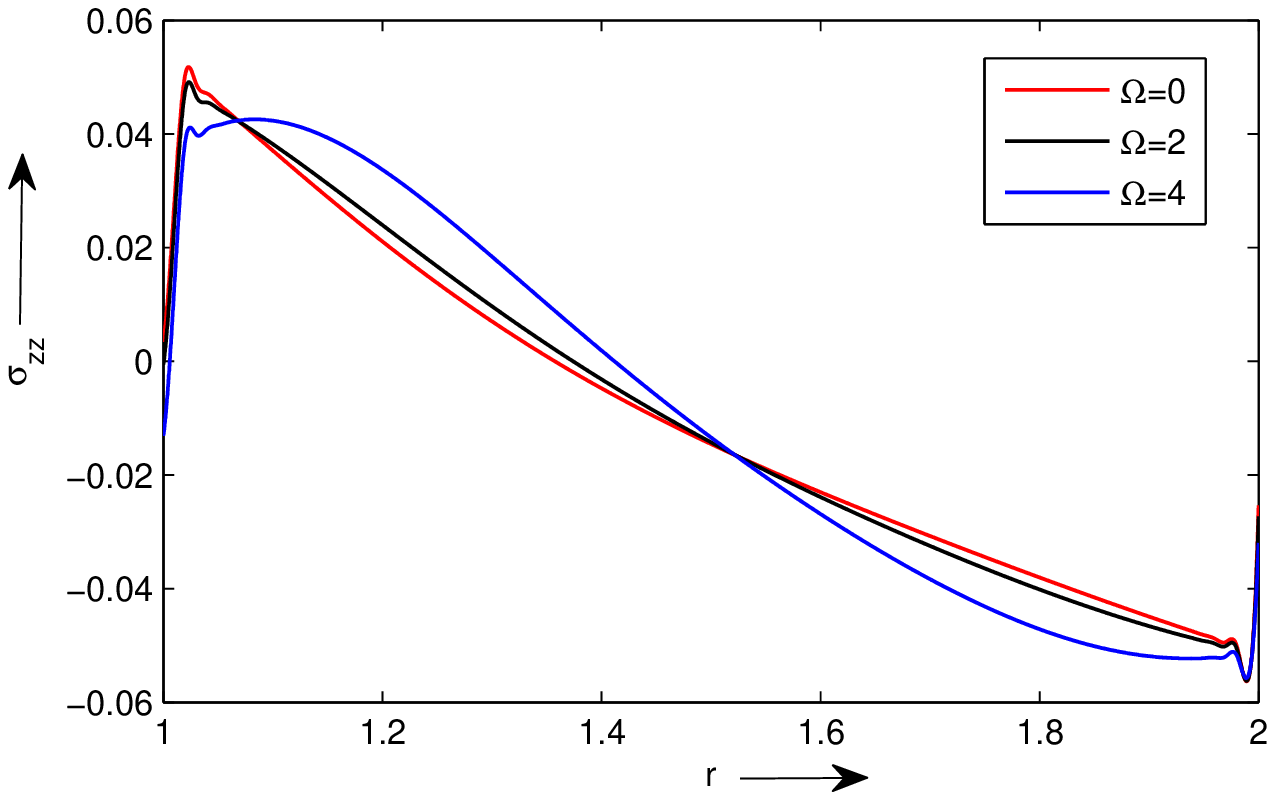}\\
Fig.7~~~ Distribution of stress($\sigma_{zz}$) vs. $r$ for $m = - 0.5$ and fixed values of $\Omega$ in G-L theory\\
\vspace*{1.5cm}
\includegraphics[width=4.8in,height=3.5in ]{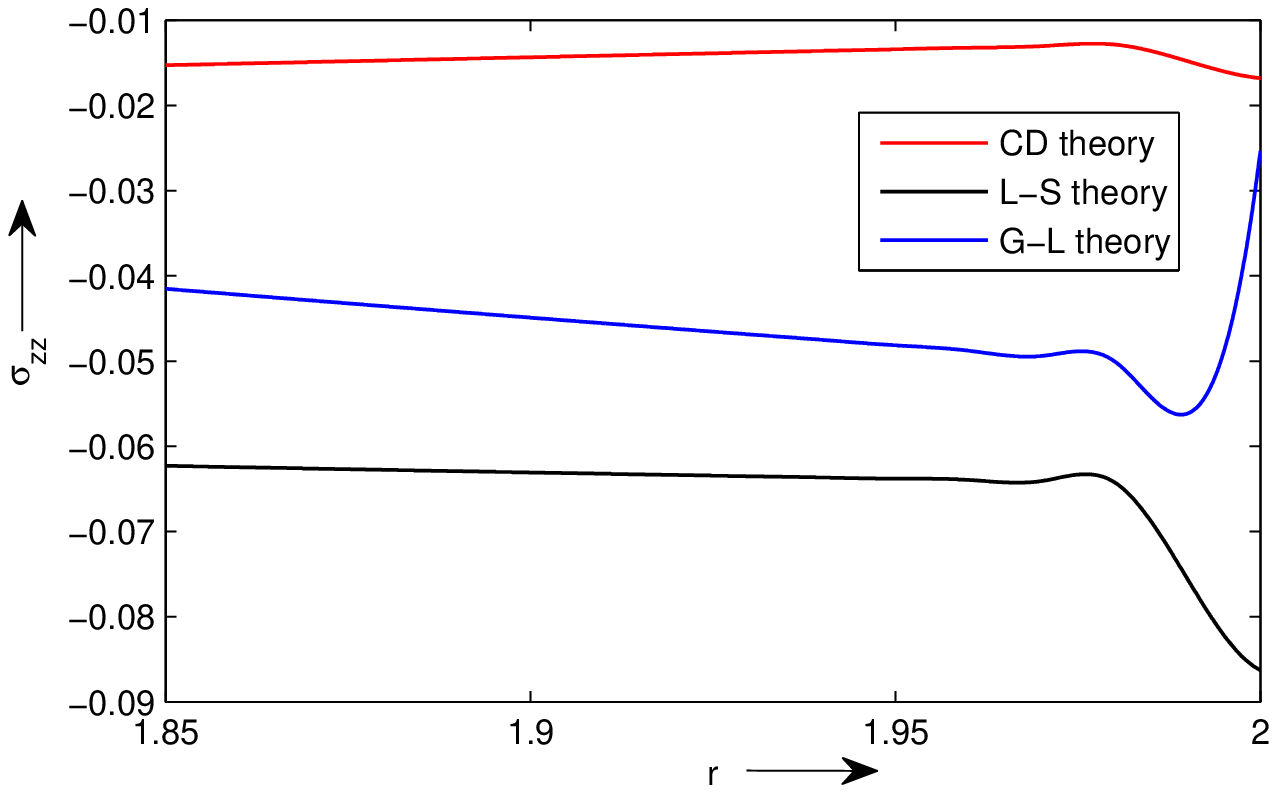}\\
Fig.8~~~ Distribution of stress($\sigma_{zz}$) vs. $r$ for $m = - 0.5$ and $\Omega=0$ in three theories\\
\end{center}
\end{minipage}
\vspace*{.5cm}\\

\textheight 22.0cm 
\begin{minipage}{1.0\textwidth}
  \begin{center}
\includegraphics[width=4.8in,height=3.5in ]{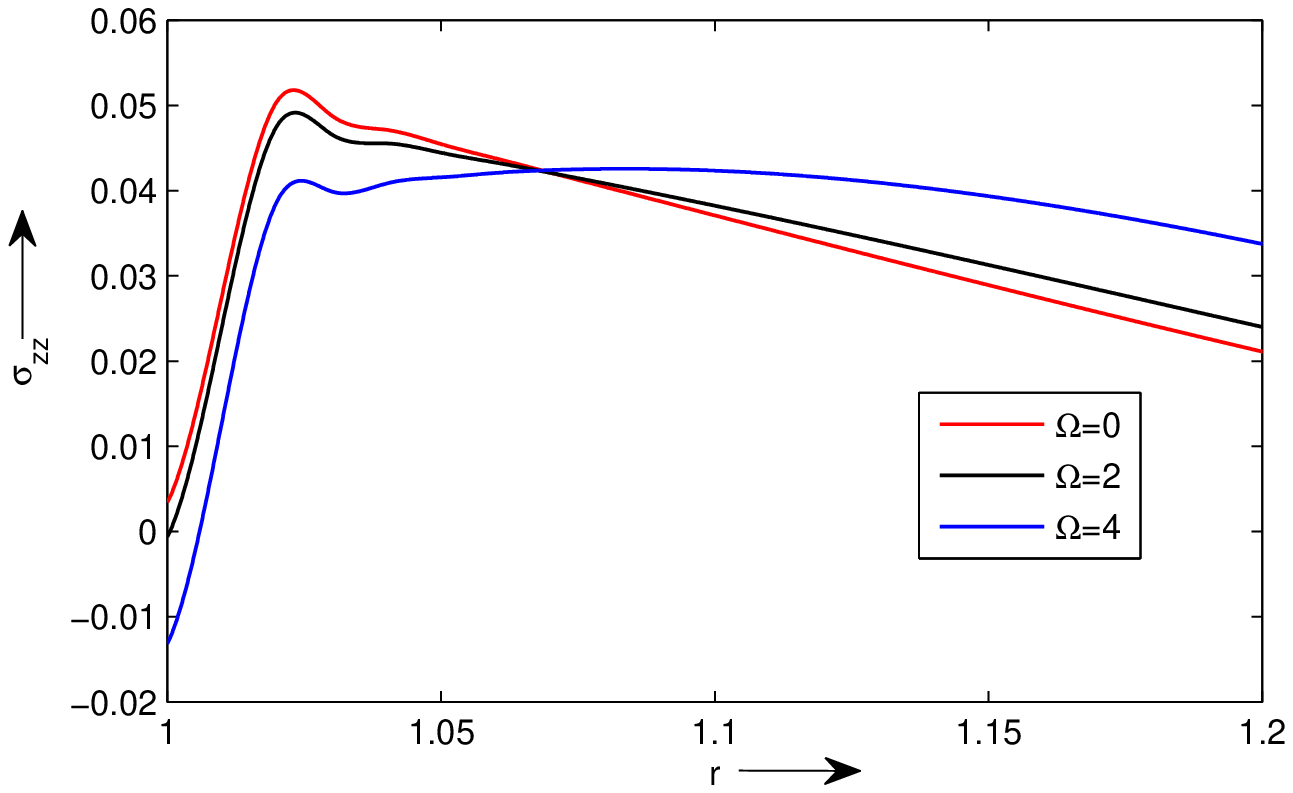}\\
Fig.9~~~ Distribution of stress($\sigma_{zz}$) vs. $r$ for $m = - 0.5$ and fixed values of $\Omega$ in G-L theory\\
\end{center}
\end{minipage}
\vspace*{.5cm}\\

Figs. 7-9 give the distribution of $\sigma_{zz}$ with $r$ for different values of parameters used in this problem. Fig.7 illustrates the distribution of $\sigma_{zz}$ vs. $r$ for $m=-0.5$ and three fixed values of $\Omega=0,2,4$ in G-L theory. This distribution is the same as of the Fig.6. Fig.8 demonstrates the variation of $\sigma_{zz}$ for $m=-0.5$ and $\Omega=0$ in three theories i.e., C-D, L-S and G-L in the region $1.85\leq r\leq 2.0$. In both these three theories, the graph of $\sigma_{zz}$ is almost parallel in the region $1.85\leq r\leq 1.97$ and also in the region $1.97\leq r \leq 2.0$, the graph of $\sigma_{zz}$ gives the significant characteristics. Fig.9 examines the distribution of $\sigma_{zz}$ for $m=-0.5$ and three fixed values of $\Omega=0,2,4$ in G-L theory. In the region $1\leq r\leq 1.02$, the absolute value of $\sigma_{zz}$ sharply attains the maximum at $r=1.02$, beyond which it decreases very slowly with the increase of $r$.\\

\section{Conclusion}
The investigation of magnetothermoelastic problem for a nonhomogeneous, isotropic and rotating hollow cylinder subjected to a some particular initial and boundary conditions has been the main concern in this paper. The problem has been solved numerically by using finite difference method via Crank-Nicolson implicit scheme. Main findings of the present study are summarized as follow:\\

      1.~~~Numerical result especially graphical representations for displacement component, temperature distribution and stress components are compared in three theories i.e., CD, L-S and G-L. The parameters used in this problem are very much sensitive in G-L theory compared to the other two theories i.e., CD and L-S. Also gives the more accurate and significant results in this G-L theory.\\
     2.~~~Several figures (Fig.1,6,7,9,12) show that the dependency of the effect of rotation in the field of magnetothermoelasticity.\\
     3.~~~The figures (Fig.1-12) show that the displacement and stress components as well as temperature distribution are quite depends upon the thickness of the cylinder.\\
     4.~~~The effect of strong dependency of the nonhomogeneous parameter(m)is shown by the figures (Fig.1, 4 and 10). For assigning the value of $m=0$, the material property of the cylinder becomes homogeneous also assigning the value of $m\neq0$, it becomes the nonhomogeneous. The above said figures show the comparison in between homogeneity and nonhomogeneity of the material property.  So, this results and analysis are very much helpful in in many different branches of engineering especially in the field of material science for designing the new structure of material decreasing the stresses in order to satisfy various engineering applications.\\
     5.~~~This problem may be studied considering special cases only by taking $m=0$ and in the separate cases in CD, L-S and G-L theory assigning the values of parameters used in this problem.\\

\end{document}